# Molecular Properties of Butan-1-ol With Acetic Acid: A Dielectric study

Baliram G.Lone[1*] and Prakash W. Khirade [1]
[1] Department of Physics, Dr. Babasaheb Ambedkar Marathwada University,
Aurangabad (Maharashtra)-431 004(India)
Suresh C. Mehrotra [2]
[2] Department of Computer Science and I.T., Dr. Babasaheb Ambedkar Marathwada University, Aurangabad-431004(India
*Corresponding authour: Nanomaterials Research Laboratory, Department of Physics
Vinayakrao Patil Mahavidyalaya, Vaijapur, 423701.Dist. Aurangabad-Maharashtra, India
E-mail: baliram.lone@aggiemail.usu.edu

The dielectric properties of Butan-1-ol with acetic acid mixture have been studied in the frequency range10 MHz to 20 GHz using time domain spectroscopy in the reflection mode, at various temperatures (i.e. on 288 K, 298 K, 308 K and 318 K) and at eleven different concentrations. Dielectric parameters such as static dielectric constant and relaxation time have been determined. Excess dielectric constant, Excess inverse relaxation time, Kirkwood correlation factor and the thermodynamic parameters ($\Delta H$ & $\Delta S$) have been estimated using these dielectric parameters.
   A drastic change in molecular structure of Butan-1-ol with acetic acid mixture observed due to strong intermolecular association among the molecules of complex sysytem. Thermodynamic parameters explored to insight molecular structures changes of mixtures at different temperatures.

## 1. Introduction

The study of mixtures of normal alcohols and carboxylic acids have been a subject of interest to several workers[1-5] because they exhibit very strong association due to strong hydrogen bonding between –OH group of primary alcohol and –COOH group of carboxylic acid [6-7].T.Fonseca and et al [8] have reported that formic acid and propionic acid each form complexes through hydrogen bonding with primary alcohols at room temperature. J. G. Kirkwood [9] has observed dipolar interaction in the mixtures of primary alcohols and carboxylic acid. J.P.Powles [10] suggested that the strong intermolecular association takes place due to strong hydrogen bonding between –OH and –COOH groups in the liquid mixture of normal alcohols and carboxylic acid, H.Wiengartner [11] also found complex formation or intermolecular association between mixture of phenols and carboxylic acid in pure state on the basis of dielectric constant, density, refractive index measurements at various temperatures. U.Kaatze and et al [12, 13] investigated that dielectric properties of carboxylic acid and water mixtures found significant information regarding inter- association in the mixture.

   The aim of the present work is to study the intermolecular interaction and molecular behavior as a function of concentration and temperature. The system under investigation is butan-1-ol and acetic acid binary mixture. The obtained results are briefly discussed in the present work.

# EXPERIMENTAL

## 2.1 Chemicals

The Butan-1-ol, Acetic acid with 99.9%purity (Spectroscopic grade, Spectrochem Pvt.Ltd, Mumbai, India) were obtained commercially and used without further purification. The solutions were prepared at room temperature with eleven different volume percentages of acetic acid from 0 % to 100 % in steps of 10% within accuracy 0.001% error limit.

## 2.2 Apparatus

The complex permittivity spectra were studied using Time Domain Reflectometry [14-15] method. A Hewlett Packard HP 54750 sampling oscilloscope with HP 54754A TDR plug-in module has been used. A fast rising step voltage pulse of about 40 ps rise time generated by a tunnel diode was propagated through a flexible coaxial cable. The sample was placed at the end of the coaxial line in the standard military application (SMA) coaxial cell of 3.5 mm outer diameter and 1.35 mm effective pin length. The change in the pulse after reflection from the sample placed in the cell was monitored by the sampling oscilloscope. In this experiment, a time window of 5 ns was used. The reflected pulses without sample, $R_1(t)$, and with sample, $R_x(t)$, were digitized in 1024 points in the memory of the oscilloscope and transferred to a PC through a 1.44 MB floppy diskette drive.

The temperature controller system with water bath and a thermostat was used to maintain constant temperature within the accuracy limit of ±1°C. The sample cell was surrounded by a heat insulating container through which water at constant temperature, is circulated using a temperature controller system.

## 3. Data analysis

The time dependent data were processed to obtain complex reflection coefficient spectra $\rho^*(\omega)$ over the frequency from 10 MHz to 20 GHZ using Fourier transformation [16, 17] as

$$\rho^*(\omega) = \left[\frac{c}{j\omega d}\right] \cdot \left[\frac{p(\omega)}{q(\omega)}\right] \qquad (1)$$

where $p(\omega)$ and $q(\omega)$ are Fourier transforms of $[R_1(t) - R_x(t)]$ and $[R_1(t) + R_x(t)]$ respectively, c is speed of light, $\omega$ is angular frequency, d is effective pin length and $j = \sqrt{-1}$

The complex permittivity spectra $\varepsilon^*(\omega)$ were obtained from reflection coefficient spectra $\rho^*(\omega)$ by using bilinear calibration method [18].

The experimental values of $\varepsilon^*$ are fitted with the Debye expression [14]

$$\varepsilon(\omega)^* = \varepsilon_\infty + \frac{\varepsilon_s - \varepsilon_\infty}{1 + j\omega\tau} \qquad (2)$$

with $\varepsilon_s, \varepsilon_\infty$ and $\tau$ as fitting parameters. In equation (2), $\varepsilon_s$ is the static dielectric constant, $\varepsilon_\infty$ is the high frequency dielectric constant, $\omega$ is the angular frequency and $\tau$ is the relaxation time of the system. A nonlinear least-squares fit method was used to determine the values of these parameters.

The information about the interaction of butan-1-ol and acetic acid may be obtained from excess properties [18, 19] such as excess permittivity and excess inverse relaxation times in the mixture. The excess permittivity ($\varepsilon_s^E$) is calculated using the relation

$$\varepsilon_s^E = (\varepsilon_s)_m - [(\varepsilon_s)_A \, x_A + (\varepsilon_s)_B \, x_B] \qquad (3)$$

where x is volume fraction and suffices m, A and B represent the mixture, Butan-1-ol (solvent) and acetic acid (solute) in the mixtures respectively.

The excess static dielectric constant provides qualitative information regarding interactions in the mixture as stated below.
  a) $\varepsilon_s^E = 0$ indicates that the solute and solvent do not interact at all.
  b) $\varepsilon_s^E < 0$ indicates that the solute and solvent interact in such a way that total no. of dipoles decreases in the mixture, and hence there is decrease in net effective dipole moment.
  c) $\varepsilon_s^E > 0$ indicates that the solute and solvent interact in such a way that effective dipole moment increases.

Similarly, the excess inverse relaxation time is calculated using the relation

$$\left(\frac{1}{\tau}\right)^E = \left(\frac{1}{\tau}\right)_m - \left[\left(\frac{1}{\tau}\right)_A X_A + \left(\frac{1}{\tau}\right)_B X_B\right] \tag{4}$$

Where $(1/\tau)^E$ is excess inverse relaxation time, which indicates the average broadening of the dielectric spectra [18]. From the investigation of excess inverse relaxation time about the dynamics of butan-1-ol-acetic acid mixture interaction the excess properties listed as follows

  I. $\left(\frac{1}{\tau}\right)^E = 0$ indicates that the solute and solvent do not interact at all.

  II. $\left(\frac{1}{\tau}\right)^E < 0$ indicates that the solute-solvent interaction takes place in such a way that the effective dipoles rotate slowly.

  III. $\left(\frac{1}{\tau}\right)^E > 0$ indicates that the solute-solvent interaction produces a field due to which the effective dipoles rotate rapidly.

The static dielectric constant of the mixtures is related to Bruggeman mixture formula with volume fraction $X_B$ of acetic acid which indicates the of interaction between butan-1-ol and acetic acid. The Bruggeman factor $f_B$ is given by

$$f_B = \left(\frac{(\varepsilon_s)_m - (\varepsilon_s)_B}{(\varepsilon_s)_A - (\varepsilon_s)_B}\right) \cdot \left(\frac{(\varepsilon_s)_A}{(\varepsilon_s)_m}\right)^{\frac{1}{3}} = (1 - X_B) \tag{5}$$

According to the Eyring rate equation [20] the relaxation time $\tau$ is given by

$$\tau = \frac{h}{kT} \exp\left(\frac{\Delta H - T\Delta S}{RT}\right), \tag{6}$$

where R is the gas constant, T is the absolute temperature, k is the Boltzman constant, h is Planck's constant, $\tau$ is experimentally determined value of relaxation time and $\Delta H$, $\Delta S$ are the molar enthalpy of activation and the molar entropy of activation.

The information regarding the correlation of electric dipoles in polar liquids is obtained from Kirkwood correlation factor g [9]

$$g = \frac{9kTM}{\mu^2 \cdot N\rho} \cdot \frac{(\varepsilon_s - \varepsilon_\infty) \cdot (2\varepsilon_s + \varepsilon_\infty)}{\varepsilon_s \cdot (\varepsilon_\infty + 2)^2} \cdot \varepsilon_o \tag{7}$$

where $\mu$ is the dipole moment, $\rho$ is the density, M is the molecular weight, k is the Boltzman constant, N is the Avogadro's number and $\varepsilon_o$ is the permittivity of free space.

Modified forms of this equation have been used to study the orientation of electric dipoles in binary mixtures[21,22,23]. Two such equations used are as follows:

$$g^{eff} = \frac{9kT}{N} \cdot \frac{(\varepsilon_s - \varepsilon_\infty) \cdot (2\varepsilon_s + \varepsilon_\infty)}{\varepsilon_s \cdot (\varepsilon_\infty + 2)^2} \cdot \varepsilon_o \cdot \left(\frac{M_A}{\mu_A^2 \cdot \rho_A \cdot x_A} + \frac{M_B}{\mu_B^2 \cdot \rho_B \cdot x_B}\right) \tag{8}$$

where $g^{eff}$ is the Kirkwood correlation factor for binary mixture. $g^{eff}$ varies between $g_A$ and $g_B$, and

$$g_f = \frac{9kT}{N} \cdot \frac{(\varepsilon_s - \varepsilon_\infty) \cdot (2\varepsilon_s + \varepsilon_\infty)}{\varepsilon_s \cdot (\varepsilon_\infty + 2)^2} \cdot \varepsilon_o \cdot \left( \frac{M_A}{\mu_A^2 \cdot \rho_A \cdot x_A \cdot g_A} + \frac{M_B}{\mu_B^2 \cdot \rho_B \cdot x_B \cdot g_B} \right) \quad (9)$$

$g_A$ and $g_B$ are assumed to be affected by an amount $g_f$ in the mixture, $g_f = 1$ for an ideal mixture, and deviation from unity indicates interaction between two components of the mixture.

## 4. RESULTS AND DISCUSSION

The dielectric constant ($\varepsilon_s$) and relaxation time ($\tau$) are obtained by fitting experimental data with the Debye equation are listed in table 1 and table 2, respectively. It is observed that the dielectric constant ($\varepsilon_s$) increases with increase in percentage of butan-1-ol and it decreases with increase in temperature. The relaxation time for the butan-1-ol - acetic acid system increases with increase in concentration of butan-1-ol. The figure 1 A and B shows the RAW spectra and corrected spectra for 50% of Butanol-Acetic acid at 298K. The plot of static dielectric constant ($\varepsilon_s$) and relaxation time ($\tau$) with volume fraction of butan-1-ol at different temperature are shown in figures 2 and 3. The nonlinear behaviour of static dielectric constant and relaxation time as seen in figures 2 and 3 suggests that the intermolecular association takes place in the system.

The behavior of excess permittivity and excess inverse relaxation time provides information regarding the arrangement of molecules in the mixtures. The excess permittivity $(\varepsilon_s)^E$ of mixtures calculated using equation 3. The variation of excess permittivity with volume fraction of butan-1-ol at 288 K, 298 K, 308 K and 318 K is shown in figure 4. The nature of the figure shows that $(\varepsilon_s)^E$ positive in acetic acid rich region and negative in butan-1-ol rich region. Negative $(\varepsilon_s)^E$ represent that the molecules of the mixtures may form multimers due to association between –OH and -COOH bonding in such a way that the effective dipole gets reduced. This association of butan-1-ol - acetic acid through hydrogen bonding helps in the formation of multimer like structures. Positive $(\varepsilon_s)^E$ represent that molecules of the mixtures may form monomers or dimmers structures in such a way that the number of effective dipoles increases.

The variation of $(1/\tau)^E$ with volume fraction of butan-1-ol calculated using equation (4) at 288 K, 298 K, 308 K and 318 K is shown in fig.5. The nature of plot confirms that the $(1/\tau)^E$ is negative from 0.1 to 0.5 volume fraction of butan-1-ol, which indicates that the effective dipoles rotate slowly. The $(1/\tau)^E$ is positive from 0.6 to 1 volume fraction of butan-1-ol, which confirms that in this region molecular interaction takes place in such a way that the dipoles rotate rapidly.

The plot of Bruggeman factor $f_B$ and volume fraction of butan-1-ol represents nonlinear behavior as shown in fig.6. From this plot it can be concluded that there is a strong intermolecular interaction at 0.1 and 0.9 volume fraction of butan-1-ol.

The molar enthalpy of activation ($\Delta H$) and the molar entropy of activation ($\Delta S$) has been calculated by using equation (6). The calculated values of $\Delta H$ and $\Delta S$ are listed in table 3. From table 3 we conclude that the more negative values of $\Delta S$ observed for the mixture in acetic acid rich region suggests that there is a strong H-bonding between butan-1-ol and acetic acid in the mixture. It is also supported by the low $\Delta H$ values in acetic acid rich region and high in buten-1-ol rich region. This means that the forces of intermolecular attraction are greater causing a still more highly ordered structure in acetic acid rich region. Hence the complexes formed by butan-1-ol +acetic acid mixture through H-bonding are more ordered in acetic acid rich region and less ordered in butan-1-ol rich region.

The Kirkwood correlation factor g is calculated by using equation (7) and is tabulated in table 4.a. It can be seen that the molecules prefer an ordering with antiparallel dipoles up to 0.8 volume fraction of butan-1-ol and thereafter dipoles prefer parallel orientation. The same behaviour is observed for $g^{eff}$, which are shown in table 4 b.

The $g_f$ values are other than 1 for all concentrations at all temperatures indicating interaction between the two components of the mixture, as seen from the values given in table 5. It is interesting to see that maximum deviation in $g_f$ values from unity are at 0.1 and 0.9 volume fraction of butan-1-ol, indicating strong intermolecular interaction.

Figure 7 shows the variation of estimated values of enthalpy and entropy as a function of volume fraction of ethanol with acetic acid.it interprets strong interaction between molecules of solute -solvent at 0.4 percent of mixture taking place.

Figure 8. shows the variation of estimated values of excess enthalpy and excess entropy as a function of volume fraction of ethanol with acetic acid. The studied plot indicates negative values at 0.4 and 0.8 signifies more strong interaction occurred.

## 5. Conclusion

Present investigation signifies there is change in molecular structures of Butanol-acetic acid molecules by means of intermolecular association between two different functional groups i.e. Butanol (-OH) and acetic acid (-COOH) groups. It is observed that the dielectric constant ($\varepsilon_s$) increases with increase in percentage of butan-1-ol and it decreases with increase in temperature. The relaxation time for the butan-1-ol - acetic acid system increases with increase in concentration of butan-1-ol.

The $(\varepsilon_s)^E$ positive in acetic acid rich region and negative in butan-1-ol rich region. Negative $(\varepsilon_s)^E$ represent that the molecules of the mixtures may form multimers due to association between –OH and -COOH bonding in such a way that the effective dipole gets reduced. the more negative values of $\Delta S$ observed for the mixture in acetic acid rich region suggests that there is a strong H-bonding between butan-1-ol and acetic acid in the mixture. It is also supported by the low $\Delta H$ values in acetic acid rich region and high in buten-1-ol rich region. This means that the forces of intermolecular attraction are greater causing a still more highly ordered structure in acetic acid rich region.

Due to the presence of carboxylic group the effect of the –OH bond is reduced. The Butan-ol-acetic acid intermolecular association contribute to significant changes in dielectric properties of complex system.

## 6.Acknowledgement

The financial assistance from the department of science and technology (DST), New Delhi, India, is thankfully acknowledged.

Table1. The static dielectric constant ($\varepsilon_s$) for Butan-1-ol –Acetic acid binary mixture at different temperatures.

| Volume fraction of butan-1-ol | 288 K | 298 K | 308 K | 318 K |
|---|---|---|---|---|
| 1 | 17.76 | 16.73 | 16.08 | 14.47 |
| 0.9 | 15.30 | 14.70 | 12.28 | 11.31 |
| 0.8 | 14.22 | 13.21 | 11.69 | 11.28 |
| 0.7 | 13.91 | 13.15 | 11.19 | 10.29 |
| 0.6 | 13.34 | 12.74 | 11.13 | 9.32 |
| 0.5 | 12.33 | 11.63 | 9.91 | 9.18 |
| 0.4 | 11.71 | 11.06 | 9.80 | 8.16 |
| 0.3 | 10.86 | 10.23 | 8.62 | 7.94 |
| 0.2 | 10.71 | 10.00 | 8.39 | 7.25 |
| 0.1 | 09.84 | 9.49 | 7.72 | 6.80 |
| 0 | 6.76 | 6.02 | 5.58 | 5.11 |

Table 2. The relaxation time ($\tau$) for Butan-1-ol –Acetic acid binary mixture at different temperatures.

| Volume fraction of butan-1-ol | 288 K | 298 K | 308 K | 318 K |
|---|---|---|---|---|
| 1 | 497.70 | 431.84 | 326.46 | 265.64 |
| 0.9 | 246.06 | 218.91 | 168.76 | 145.55 |
| 0.8 | 224.80 | 205.71 | 157.97 | 137.51 |
| 0.7 | 200.54 | 181.69 | 146.98 | 123.53 |
| 0.6 | 190.75 | 166.12 | 142.93 | 113.11 |
| 0.5 | 173.23 | 156.03 | 127.96 | 105.29 |
| 0.4 | 160.34 | 139.84 | 129.04 | 92.94 |
| 0.3 | 161.77 | 139.52 | 116.57 | 98.75 |
| 0.2 | 140.95 | 129.78 | 111.89 | 94.85 |
| 0.1 | 135.65 | 114.56 | 105.37 | 90.16 |
| 0 | 108.82 | 95.42 | 78.89 | 69.19 |

Table3 ΔH and ΔS for mixtures of butan-1-ol+acetic acid binary mixture.

| Volume fraction of butan-1-ol | ΔH (KJ/mole) | ΔS (KJ/mole) |
|---|---|---|
| 1 | 13.914 | -0.018 |
| 0.9 | 11.435 | -0.021 |
| 0.8 | 10.689 | -0.023 |
| 0.7 | 10.121 | -0.024 |
| 0.6 | 10.512 | -0.022 |
| 0.5 | 10.315 | -0.022 |
| 0.4 | 10.452 | -0.021 |
| 0.3 | 10.111 | -0.022 |
| 0.2 | 7.615 | -0.030 |
| 0.1 | 7.454 | -0.030 |
| 0 | 9.270 | -0.022 |

Table 4 a. The Kirkwood correlation factor g for butan-1-ol+acetic acid at different temperatures.

| Volume fraction of butan-1-ol | 288 K | 298 K | 308 K | 318 K |
|---|---|---|---|---|
| 1 | 0.366 | 0.354 | 0.350 | 0.320 |
| 0.9 | 0.326 | 0.322 | 0.269 | 0.251 |
| 0.8 | 0.321 | 0.304 | 0.271 | 0.268 |
| 0.7 | 0.339 | 0.328 | 0.279 | 0.259 |
| 0.6 | 0.358 | 0.350 | 0.307 | 0.252 |
| 0.5 | 0.370 | 0.357 | 0.301 | 0.281 |
| 0.4 | 0.411 | 0.396 | 0.351 | 0.282 |
| 0.3 | 0.468 | 0.449 | 0.369 | 0.340 |
| 0.2 | 0.639 | 0.605 | 0.494 | 0.412 |
| 0.1 | 1.002 | 0.998 | 0.766 | 0.651 |
| 0 | 3.306 | 2.809 | 2.519 | 2.168 |

Table 4.b The Kirkwood correlation factor $g^{eff}$ for butan-1-ol+acetic acid at different temperatures.

| Volume fraction of butan-1-ol | 288 K | 298 K | 308 K | 318 K |
|---|---|---|---|---|
| 1 | 0.366 | 0.354 | 0.350 | 0.320 |
| 0.9 | 0.342 | 0.338 | 0.282 | 0.263 |
| 0.8 | 0.352 | 0.334 | 0.298 | 0.294 |
| 0.7 | 0.390 | 0.377 | 0.320 | 0.297 |
| 0.6 | 0.430 | 0.421 | 0.368 | 0.303 |
| 0.5 | 0.464 | 0.447 | 0.377 | 0.352 |
| 0.4 | 0.536 | 0.517 | 0.458 | 0.368 |
| 0.3 | 0.635 | 0.609 | 0.501 | 0.460 |
| 0.2 | 0.892 | 0.845 | 0.690 | 0.575 |
| 0.1 | 1.402 | 1.383 | 1.072 | 0.911 |
| 0 | 3.306 | 2.809 | 2.519 | 2.168 |

Table 5 The Kirkwood correlation factor $g_f$ for butan-1-ol+acetic acid at different temperatures.

| Volume fraction of butan-1-ol | 288 K | 298 K | 308 K | 318 K |
|---|---|---|---|---|
| 1 | 288 K | 298 K | 308 K | 318 K |
| 0.9 | 1.000 | 1.000 | 1 | 1 |
| 0.8 | 0.909 | 0.931 | 0.789 | 0.807 |
| 0.7 | 0.904 | 0.894 | 0.811 | 0.879 |
| 0.6 | 0.961 | 0.974 | 0.845 | 0.863 |
| 0.5 | 1.005 | 1.038 | 0.933 | 0.845 |
| 0.4 | 1.016 | 1.040 | 0.905 | 0.934 |
| 0.3 | 1.073 | 1.110 | 1.020 | 0.912 |
| 0.2 | 1.116 | 1.163 | 1.002 | 1.029 |
| 0.1 | 1.278 | 1.339 | 1.160 | 1.088 |
| 0 | 1.374 | 1.537 | 1.288 | 1.246 |
|  | 1.000 | 1.000 | 1 | 1 |

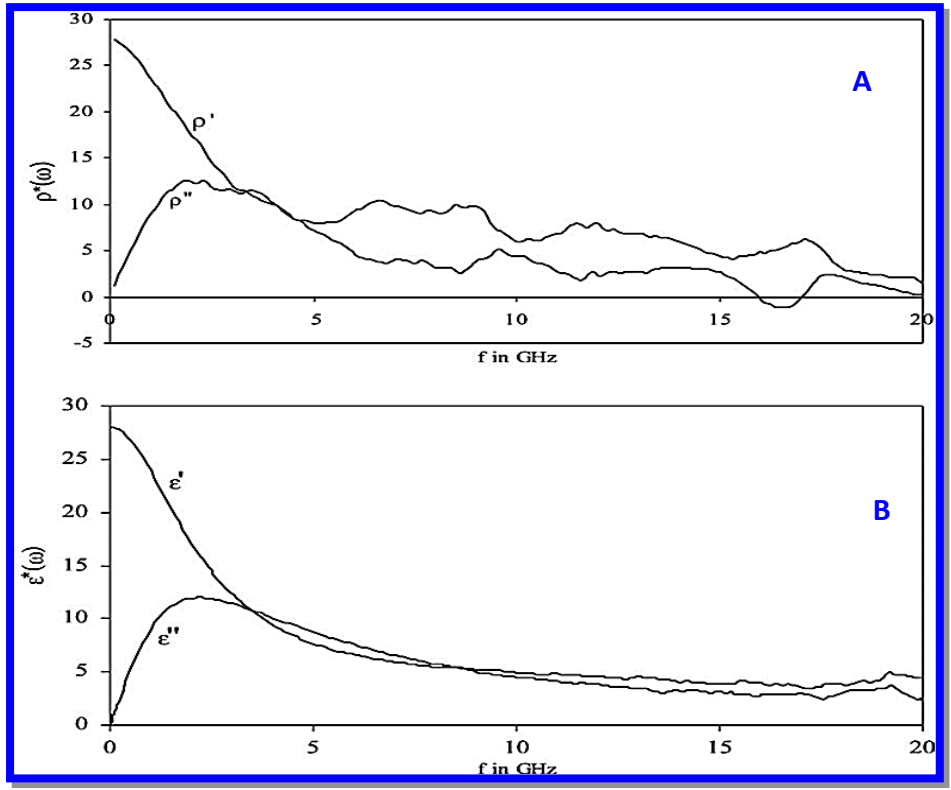

**Figure 1.** (A) ρ∗(ω) spectra for 50% Butanol- acetic Acid at 298 K. (B) ε∗(ω) spectra for 50% Butanol- acetic Acid at 298 K.

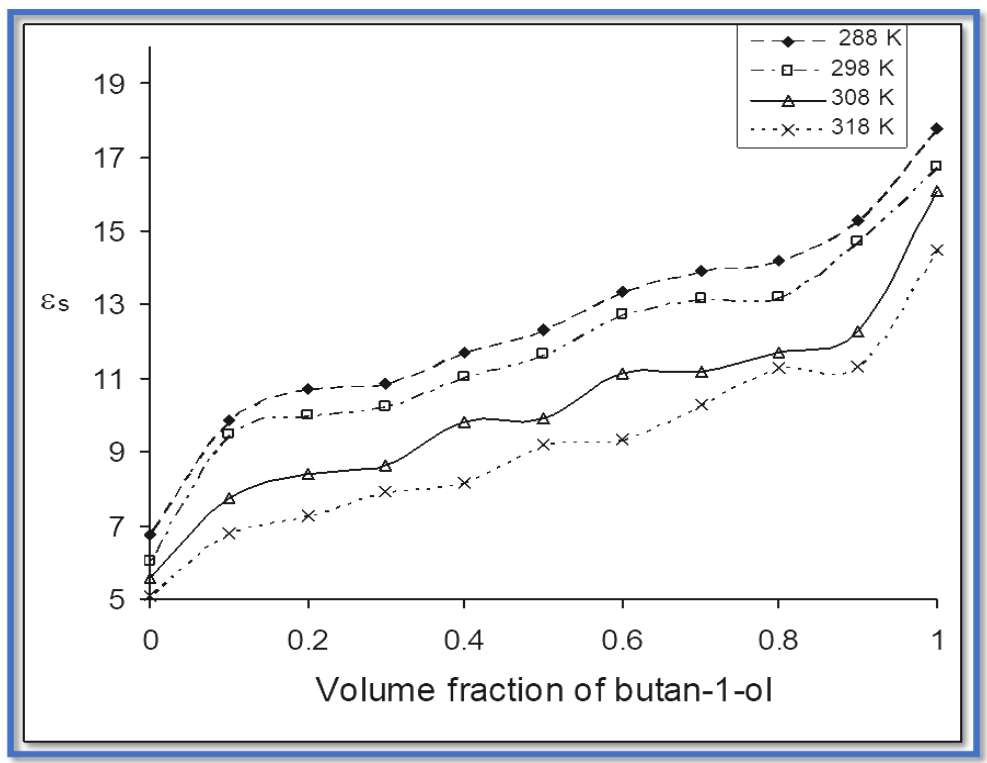

Fig.2 The static dielectric constant ($\varepsilon_s$) versus volume fraction of Acetic acid in Butan-1-ol at different temperatures.

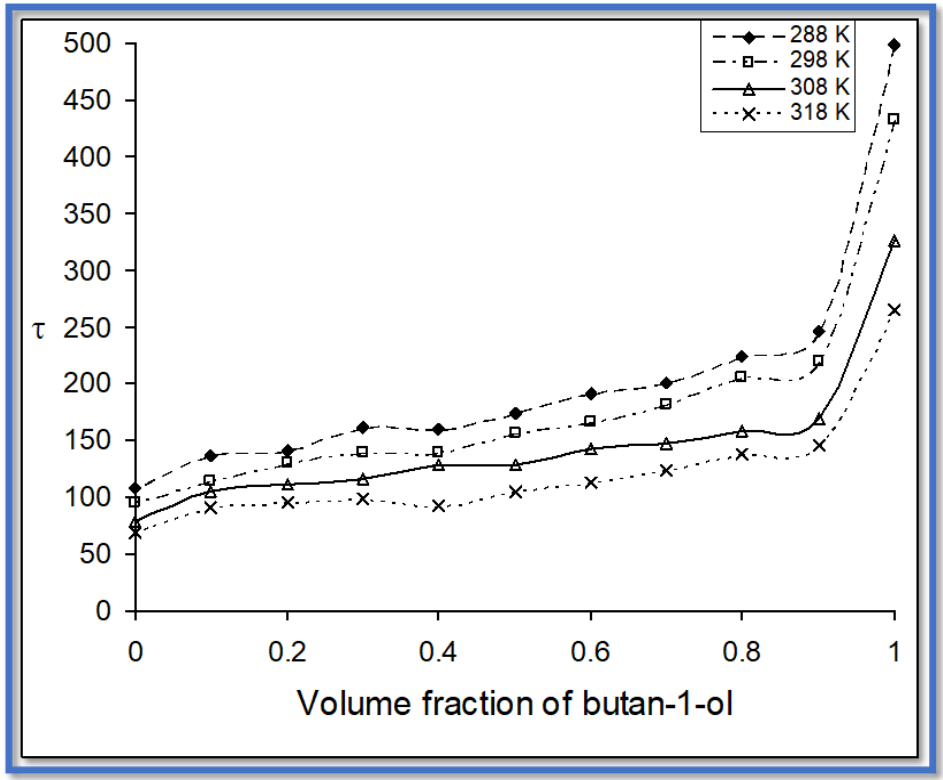

Fig.3 The relaxation time (τ) versus volume fraction of Acetic acid in Butan-1-ol different temperatures.

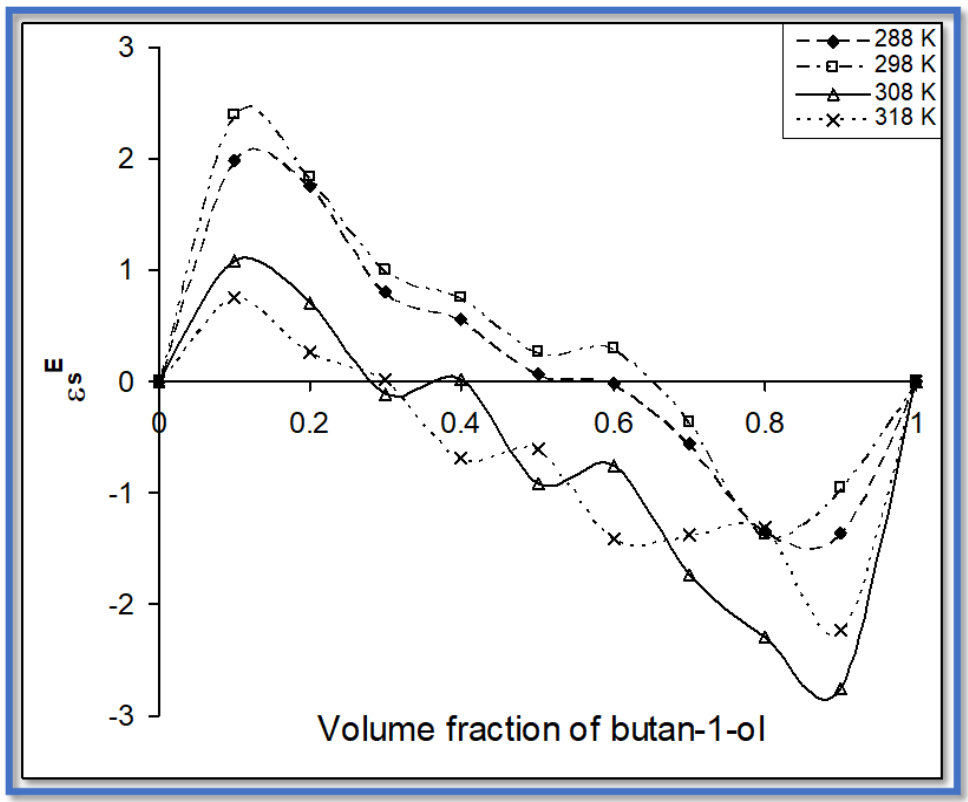

Fig.4. The excess dielectric constant ($\varepsilon_s^E$) versus volume fraction of Acetic acid in Butan-1-ol at different temperatures.

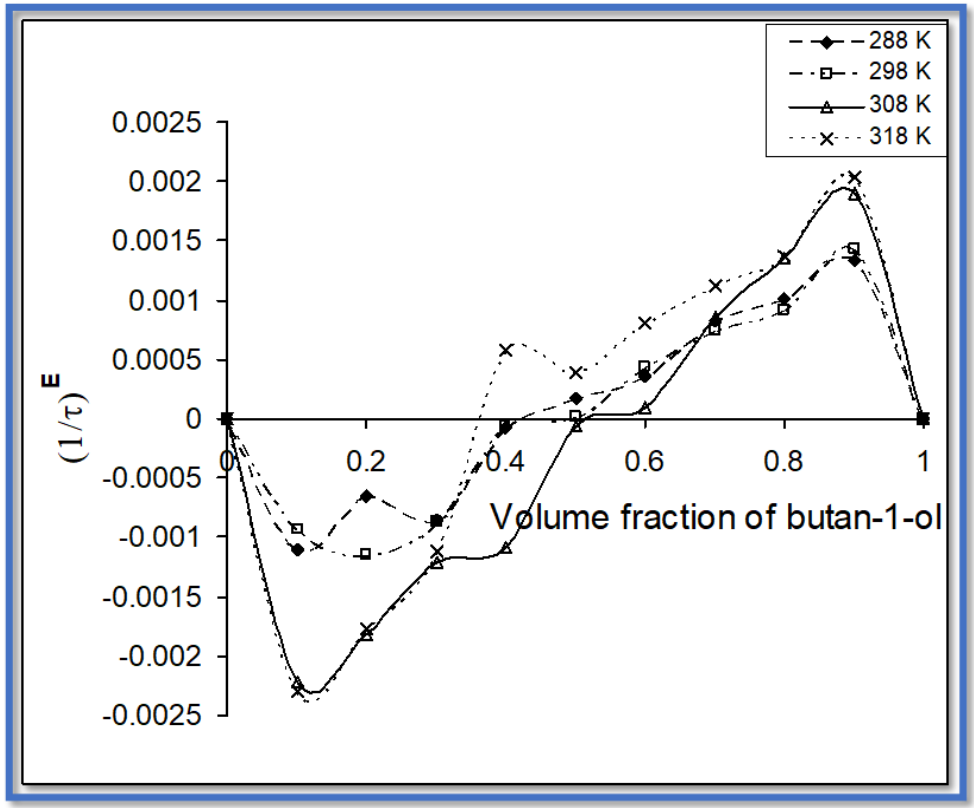

Fig.5. The inverse relaxation time $(1/\tau)^E$ versus volume fraction of Acetic acid in Butan-1-ol different temperatures.

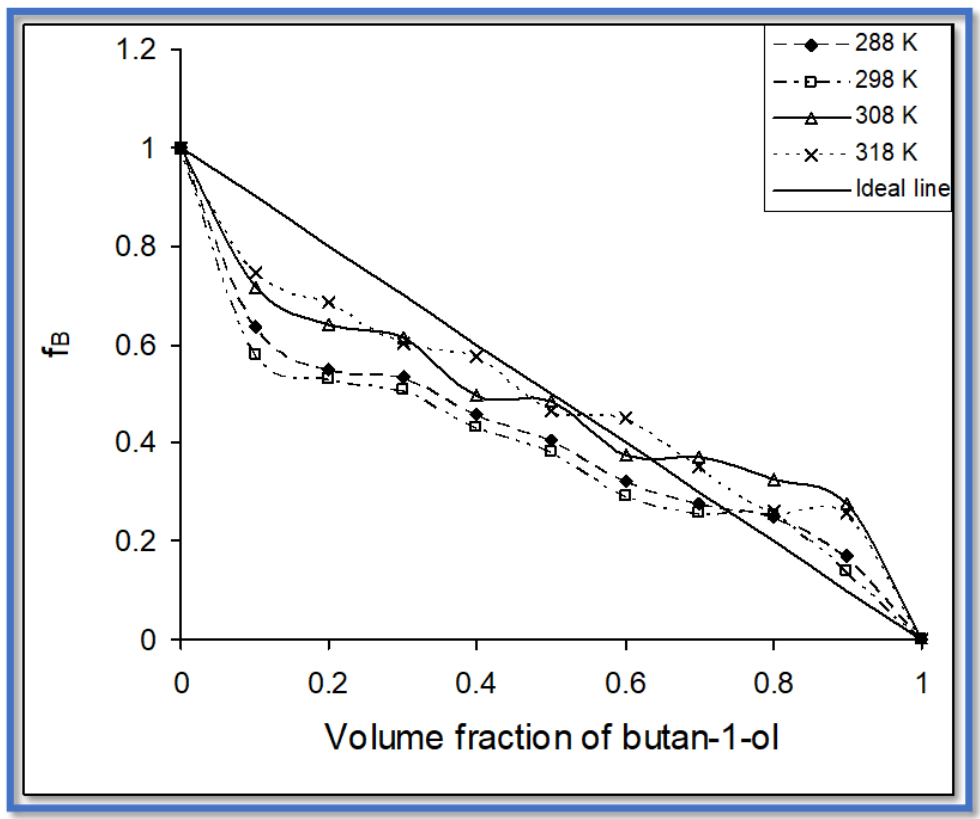

Fig.6 The Bruggeman plot for butan-1-ol-acetic acid at different temperatures.

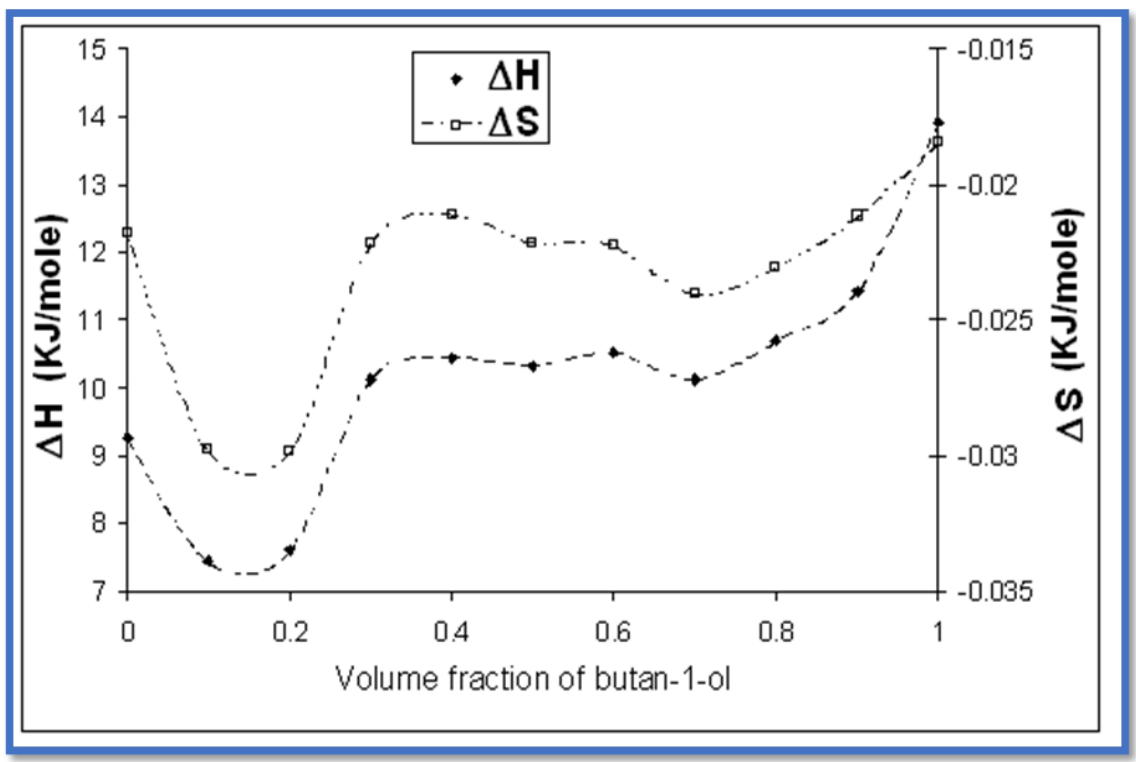

Fig.7 The plot of the molar enthalpy of activation ΔH and molar entropy of activation ΔS for mixtures of butan-1-ol-acetic acid at different temperatures.

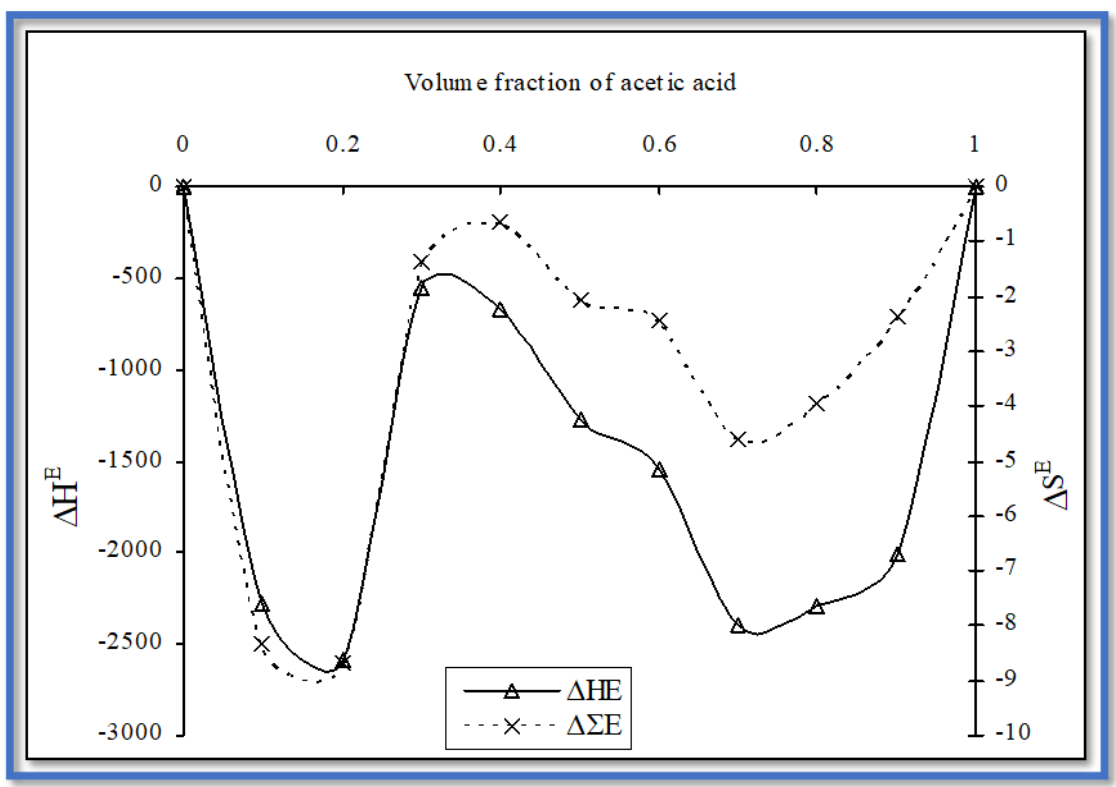

Fig.8. The plot of an excess molar enthalpy of activation $(\Delta H)^E$ and an excess molar entropy of activation $(\Delta S)^E$ for mixtures of butan-1-ol+acetic acid binary mixture at different temperatures.

**Captions to tables and figures**

Table 1. Estimated static dielectric constant for butan-1-ol - acetic acid binary mixture at different temperatures.
Table 2. Estimated relaxation time for buta1-ol - acetic acid binary mixture at different temperatures.
Table 3. Estimated values of enthalpy ($\Delta H$) and entropy ($\Delta S$)
Table 4 a. The Kirkwood correlation factor g for butan-1-ol+acetic acid at different temperatures.
Table 4.b The Kirkwood correlation factor $g^{eff}$ for butan-1-ol+acetic acid at different temperatures.
Table 5 The Kirkwood correlation factor $g_f$ for butan-1-ol+acetic acid at different temperatures.

Figure 1. (A) $\rho*(\omega)$ spectra for 50% Butanol- acetic Acid at 298 K. (B) $\varepsilon*(\omega)$ spectra for 50% Butanol- acetic Acid at 298 K.
Figure 2. Variation of estimated static dielectric constant ($\varepsilon_s$) versus volume fraction ($x_2$) of ethanol in methanol at different temperatures.
Figure 3. Variation of estimated relaxation time ($\tau$) versus volume fraction ($x_2$) of ethanol in methanol at different temperatures.
Figure 4. Variation of estimated excess dielectric constant $\varepsilon^E$ as a function of volume fraction of ethanol in methanol at different temperatures.
Figure 5. Variation of estimated excess inverse relaxation time $(1/\tau)^E$ as a function of volume fraction of ethanol in methanol at different temperatures.
Figure 6. Variation of estimated values of Bruggeman factor as a function of volume fraction of ethanol in methanol.
Figure 7. Variation of estimated values of enthalpy and entropy as a function of volume fraction of ethanol.
Figure 8. Variation of estimated values of enthalpy and entropy as a function of volume fraction of ethanol.